\documentclass[useAMS,usenatbib,referee]{mn2e}
\usepackage{epsfig}
\usepackage{graphicx}
\usepackage[intlimits]{amsmath}
\usepackage{amssymb}

\begin{document}

\title[Hard eclipse of SS433]{
Peculiar nature of hard X-ray eclipse in SS433 from INTEGRAL observations}

\author[A.M. Cherepashchuk et al.]{A.M. Cherepashchuk$^{1}$,
R.A. Sunyaev$^{2}$,
K.A. Postnov$^{1}$, E.A. Antokhina$^{1}$,
\and S.V. Molkov$^{2,3}$ 
\thanks{E-mail: cherepashchuk@gmail.com(AMCh);
kpostnov@gmail.com(KAP); elant@sai.msu.ru(EAA)} \\
$^{1}${\sl Sternberg Astronomical Institute, Universitetskij pr. 13, Moscow 119992, Russia}\\ 
$^{2}${\sl Space Research Institute, Moscow, Russia}\\
$^{3}${\sl CESR, Toulouse, France}\\
}
\date{Accepted ......  Received ......; in original form ......
}

\maketitle

\begin{abstract}
The analysis of hard X-ray INTEGRAL observations (2003-2008) of 
superaccreting galactic microquasar SS433 at precessional
phases of the source with the maximum disk opening 
angle is carried out. It is found that the shape and width of the primary 
X-ray eclipse is strongly variable suggesting additional absorption in dense stellar wind and gas outflows from the
optical A7I-component and the wind-wind collision region. 
The independence of the observed hard X-ray spectrum on the accretion disk
precessional phase suggests that hard X-ray emission ($20-100$~keV)
is formed in an extended, hot, quasi-isothermal corona, probably 
heated by interaction of relativistic jet with 
inhomogeneous wind outflow from 
the precessing supercritical accretion disk.
A joint modeling of X-ray eclipsing 
and precessional hard X-ray variability of SS433 revealed by INTEGRAL by a 
geometrical model suggests the binary mass ratio 
$q=m_x/m_v\simeq 0.25\div 0.5$. The absolute minimum of joint
orbital and precessional $\chi^2$ residuals is reached 
at $q\simeq 0.3$.  
The found binary 
mass ratio range allows us to explain the substantial 
precessional variability of the minimum brightness at the middle
of the primary optical eclipse. 
For the mass function of
the optical star $f_v=0.268 M_\odot$ as derived from Hillwig \& Gies data, 
the obtained value of $q\simeq 0.3$ 
yields the masses of the components $m_x\simeq 5.3 M_\odot$, 
$m_v\simeq 17.7 M_\odot$, confirming the black hole nature of
the compact object in SS433. 

\end{abstract}


\section{Introduction}
\label{sec:intro}

SS433 is a massive close binary system at advanced evolutionary stage
\citep{Margon84, Cher81, Cher88, Fab04}. This unique galactic X-ray
binary with 
precessing relativistic jets ($v=0.26c$, where $c$ is the speed of light)
exhibits several variabilities, including the precessional one (with the
period 
$P_{prec}\simeq 162$~d), the eclipsing one (with the binary orbital period  
$P_{orb}\simeq 13.08$~d), and the nutational one (with the nutation period
$P_{nut}\simeq 6.28$~d) (e.g. \cite{Goransk98}, \cite{Davydov08}). SS433 is recognized 
as a galactic microquasar with precessing supercritical accretion disk
around a compact object, and has been extensively investigated 
in the optical, radio and X-ray ranges (for a comprehensive
review and references see \cite{Fab04}). Recent high-resolution optical spectroscopy of the system
\citep{hillwig08} revealed the presence of absorption lines in the
spectrum of the optical component identified as a $\sim$~A7I supergiant
star. Observed orbital Doppler shifts of the absorption lines of the optical
component and stationary HeII emission allowed Hillwig \& Gies to determine the mass ratio of the compact
($m_x$) and the optical ($m_v$) components in SS433 $q=m_x/m_v\simeq0.35$, implying the binary masses $m_x=4.3\pm0.8 M_\odot$ and $m_v=12.3\pm 3.3
M_\odot$. Similar values of $q, m_x$ and $m_v$ were obtained from the analysis of the optical
light curves of SS433 \citep{AntCher87}. 

The analysis of broad X-ray eclipses in the 1-10 keV range observed
by \textit{ASCA} and \textit{Ginga}
 satellites suggests a smaller mass ratio than that obtained from optical
spectroscopy and photometry. For example, using X-ray eclipses observed by \textit{Ginga} \cite{brinkman89} estimated $q \simeq 0.15$. In their analysis, the authors 
used the simplifying assumption that the X-ray source is compact and the 
duration of the X-ray eclipse is fully determined by the size of the
optical companion. 
\textit{Ginga} light curves at different precessional
phases presented in \cite{Kawai89} were analyzed by \cite{Ant92} using the method
of binary light curve synthesis in the model including the precessing accretion disk and jets. They found the possible range of binary mass ratios 
$q=0.15-0.25$. Using X-ray emission from SS433 
observed by the \textit{ASCA} satellite (which is mostly produced in 
precessing thermal jets), \cite{KotaniPhD} and \cite{Kotani98} 
estimated $q \simeq 0.22$. Their analysis, with account of uncertainties, 
allowed in fact a fairly broad range of binary mass ratios $q=0.06-0.31$.
Thus the average mass ratio estimate as inferred from the analysis of 
1-10 keV X-ray data is smaller ($q \le 0.25$) than that inferred from 
optical observations ($q \sim 0.35$). However, for $q < 0.25$ 
the total eclipse of the accretion disk by the optical star must occur, so 
it is hard to explain
a significant optical variability of SS433 in the primary eclipse minimum with
precessional phase, as in fact observed \citep{Goransk98}.

The INTEGRAL satellite detected   
the hard X-ray emission from SS433 up to 100 keV \citep{Cher03}. 
Our studies of SS433 based on the INTEGRAL observations
suggested the presence of a hot rarefied corona above the supercritical
accretion disk in this source \citep{Cher05, Cher06}. The
peculiar variability in the shape and width of the primary eclipse in hard
X-rays was discovered. 
The spectral analysis of archival  RXTE observations of SS433 \citep{Fil06} found evidence for strong variable X-ray absorption 
which does not support the assumption that the optical star can be considered as 
an opaque screen with sharp edges at these energies.   
This 
implies that the primary eclipse in SS433 is not purely 
geometrical and that the binary mass ratio as inferred from the analysis 
of the \textit{Ginga} and \textit{ASCA} data  
may suffer from systematic uncertainties.

The results of our previous analysis of the orbital and precessional variability in
SS433 observed by INTEGRAL \citep{Cher05, Cher06} can be summarized as
follows. The hard X-ray flux (25-50 keV) from the source clearly exhibits variability with the precessional period $P_{prec}\simeq 162.4$~d
from $\sim 3$~mCrab at the cross-over phase to $\sim 18-20$~mCrab at the
maximum disk opening phase (the $T_3$ moment, where the moving emission
lines in the SS433 spectrum are at maximum separation). 
Eclipses are observed with the orbital period $P_{orb}\simeq 13.08$~d and are very
significant. Eclipses observed close to the $T_3$ phase (at the precessional phase $\psi\simeq 0$) are the deepest ones. The hard X-ray flux (18-60~keV) at the center of the primary eclipse is still detectable at a level of $\sim 3$~mCrab, so the ratio of the maximum uneclipsed flux ($\sim 20$~mCrab) to the minimum flux value 
at the mid-eclipse is about 6-7, which is similar to the amplitude
of the precessional off-eclipse hard X-ray variability. The width of the hard X-ray eclipse is found to be larger than that in soft X-rays. The egress out of the hard X-ray eclipse is observed to be strongly variable, most probably due to absorption of the X-ray flux by accretion flows and asymmetric structured wind from the supercritical accretion disk and by the wind-wind collision region. Similar distortions of the eclipse egress were first observed by \textit{Ginga} (18.4-27.6 keV) \citep{Kawai89}. The hard X-ray spectrum (20-200 keV) does not noticeably change with  precessional phase. All these facts suggest that most of the hard X-ray flux of SS433 is generated in a hot extended corona formed in the central parts of the accretion disk.
The hot corona may be formed as a result of collision of the wind inhomogeneities with relativistic jets \citep{Begelman06}. 
 A more detailed interpretation of the broad-band (3-90 keV) X-ray continuum of SS433 in terms of the multicomponent model including the accretion disk, jet and corona has been carried out by \cite{Krivosheev08}. 

In the present paper, 
we analyze INTEGRAL observations of hard X-ray eclipses of SS433 near the T3 moment
and interpret them in terms of our multicomponent 
geometrical model with account of the peculiar shape and strong variability of the primary eclipse and precessional 
X-ray variability of the source. 
In Section \ref{s_observ} we describe 
INTEGRAL observations used for the analysis. 
In Section \ref{s_precvar} 
we discuss the precessional variability of SS433.
In Section \ref{s_hardsp} we present IBIS/ISGRI spectra
of SS433 obtained at different precessional intervals and
show that they can be fitted by one power law. 
In Section \ref{s_primecl} we study in more detail the primary 
eclipse of SS433 in hard X-rays.
Section \ref{s_geommod} depicts our 
geometrical model. In
Section \ref{s_analysis} 
the best-fit mass ratio is obtained from joint analysis of eclipsing
and precessional light curves. 
We discuss the obtained results in Section \ref{s_disc} and summarize
our findings in Section \ref{s_concl}.

\section{Observations}
\label{s_observ}

Dedicated INTEGRAL observations of 
SS433 were carried out in AO-1 for 500 ks, in AO-3 for 500 ks, in AO-4 for 466 ks, in AO-5 for $\sim 900$~ks and in AO-6 for $\sim 500$~ks. 
The AO-1 observations (INTEGRAL orbits 67-69) were performed 
at the precessional phase 0 with maximum disk opening. 
In AO-2, SS433 fell within the FOV of IBIS when observing the Sagittarius
Arm Tangent region, but no X-ray eclipses occurred during this program. 
In AO-3, SS433 was observed around different precessional phases (INTEGRAL orbits 366-369). 
One X-ray eclipse was partially measured again at the precessional phase close to zero, with an indication of a much narrower eclipse or 
a sudden mid-eclipse (at the orbital phase $\phi \sim 1.03$) flux increase. 
Based on these data, the model for the source eclipse has been 
constructed in \citep{Cher05, Cher06}; however, 
these data were not strongly constrained by precessional variability and 
allowed a broad range of the binary mass ratio $q\sim 0.1-0.5$. 
In AO-4, the source was observed in May 2007; unfortunately, due to high variability, 
the source flux was very low, so we exclude these data from the analysis of
X-ray eclipses (set II in Fig. \ref{f_alleclipses} below). Note the strong increase in 
the apparent eclipse width. 
In AO-5, two consecutive eclipses of the source near the zero precessional phase $\psi=0$
were observed in October 2007. Adding all of these observations allowed us 
to construct hard X-ray spectra of the source at different precessional phases  
(Fig. \ref{Xspectra}). However, the statistics is still insufficient to 
make orbital phase-resolved spectroscopy. 
In AO-6, the egress out of the primary eclipse of SS433 was observed in September 2008 (set V in Fig. \ref{f_alleclipses} below). 

\begin{figure}
\includegraphics[width=0.47\textwidth]{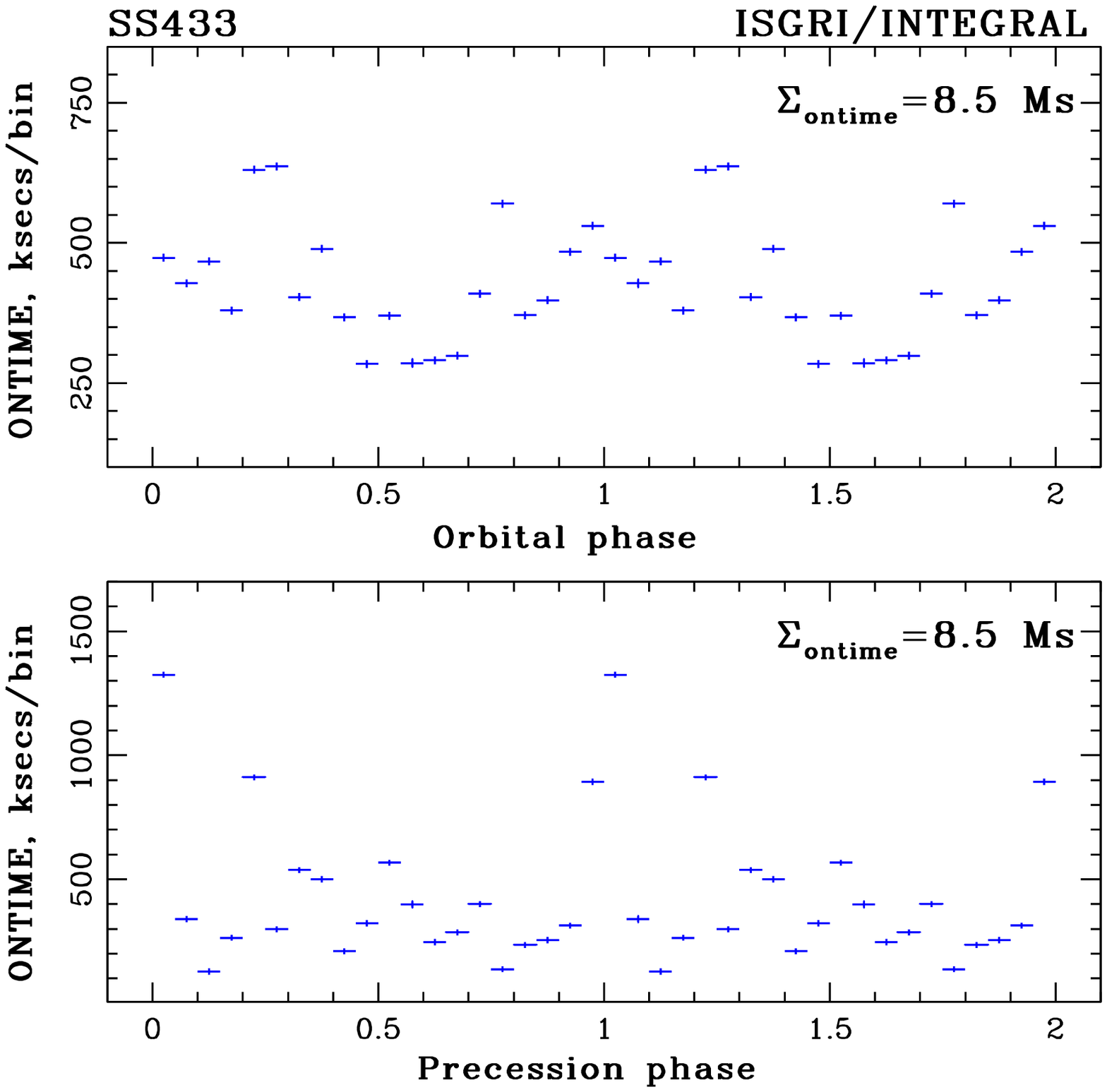}
\hfill
\includegraphics[width=0.47\textwidth]{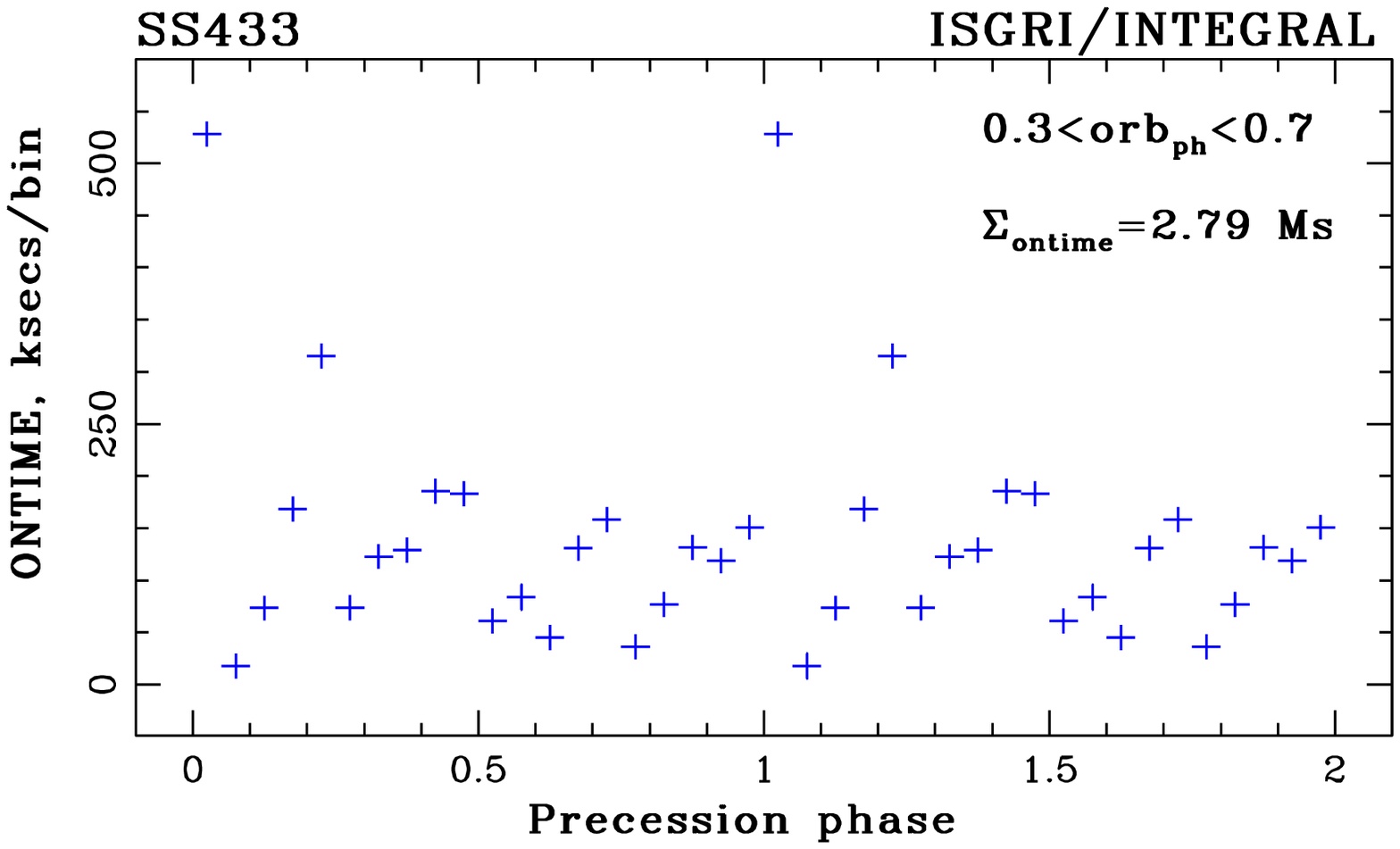}
{\caption{Left panel: distribution of the observing time 
of SS433 by the orbital (upper) and the precession (bottom)
phases. Right panel: the distribution of the duration of 
the off-eclipse data ($0.3<\psi_{orb}<0.7$) over precessional phase 
used for the construction of the  
precessional light curve.\label{distr_tot_exp}}}
\end{figure}

The INTEGRAL data were processed with both publically available
ISDC software (OSA-7 version) and the original software
package written by the IKI INTEGRAL team (for the IBIS/ISGRI telescope,
see \citep{Mol04} for more detail).
For the analysis of precessional variability of SS433 
we have used both data from our INTEGRAL observing program of
SS433 and publically available data of all observations where the
source was in the FOV of the IBIS/ISGRI telescope ($<13$ degrees). The
total exposure of the selected data is approximately $8.5$ Ms. To perform
precessional phase-resolved analysis we ascribed to each SCW (Science Window
or SCW, the natural piece of INTEGRAL data -- pointing observation with
an exposure of $\sim2-5$~ks) the appropriate orbital and precessional phases.
The phases are calculated using the ephemeris provided by \cite{Fab04}. 
The orbital primary minimum (corresponding to $\phi=0$) is 
$$
JD_{MinI} \hbox{(hel)} = 2450023.62 + 13.08211\times E\,,
$$
the zero precession phase (corresponding to the T3  moment, $\psi=0$) is
$$
JD_{T3} \hbox{(hel)}= 2443507.47 + 162.375\times E1\,.
$$
%
In Fig. \ref{distr_tot_exp} (left panels) we show the exposure distribution of
the used observations by phases. For clarity we presented two cycles.
The exposure time distribution over orbital phases is distributed more or less
homogeneously, while that over precessional phases shows an excess near the 
$T3$ moment, since the main part of our SS433 INTEGRAL
observing program was focused on observations of the primary eclipses at zero 
precessional phases where the source flux is maximal.

\subsection{Precessional variability}
\label{s_precvar}

\begin{figure}
\includegraphics[width=0.47\textwidth]{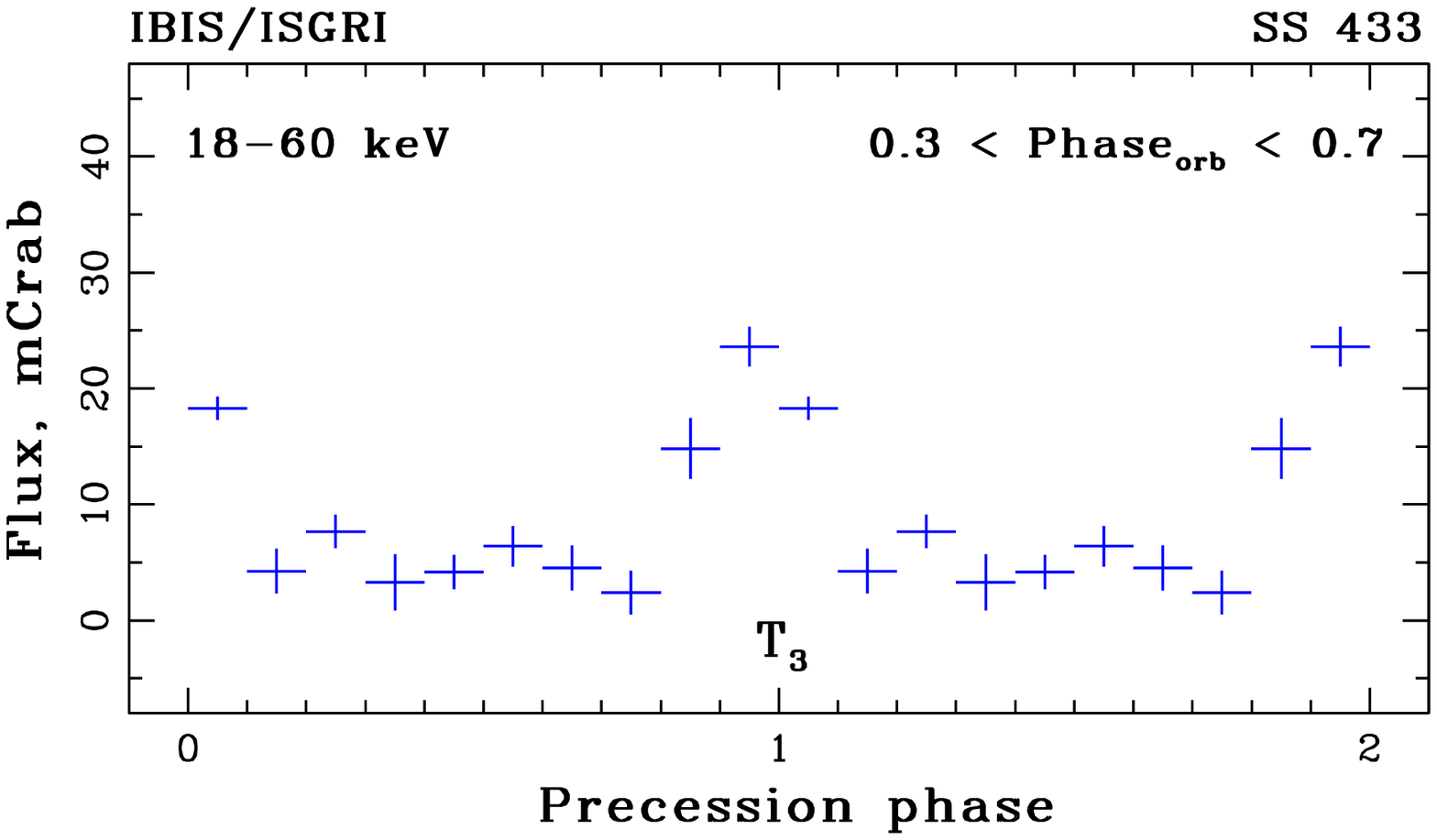}
\hfill
\includegraphics[width=0.47\textwidth]{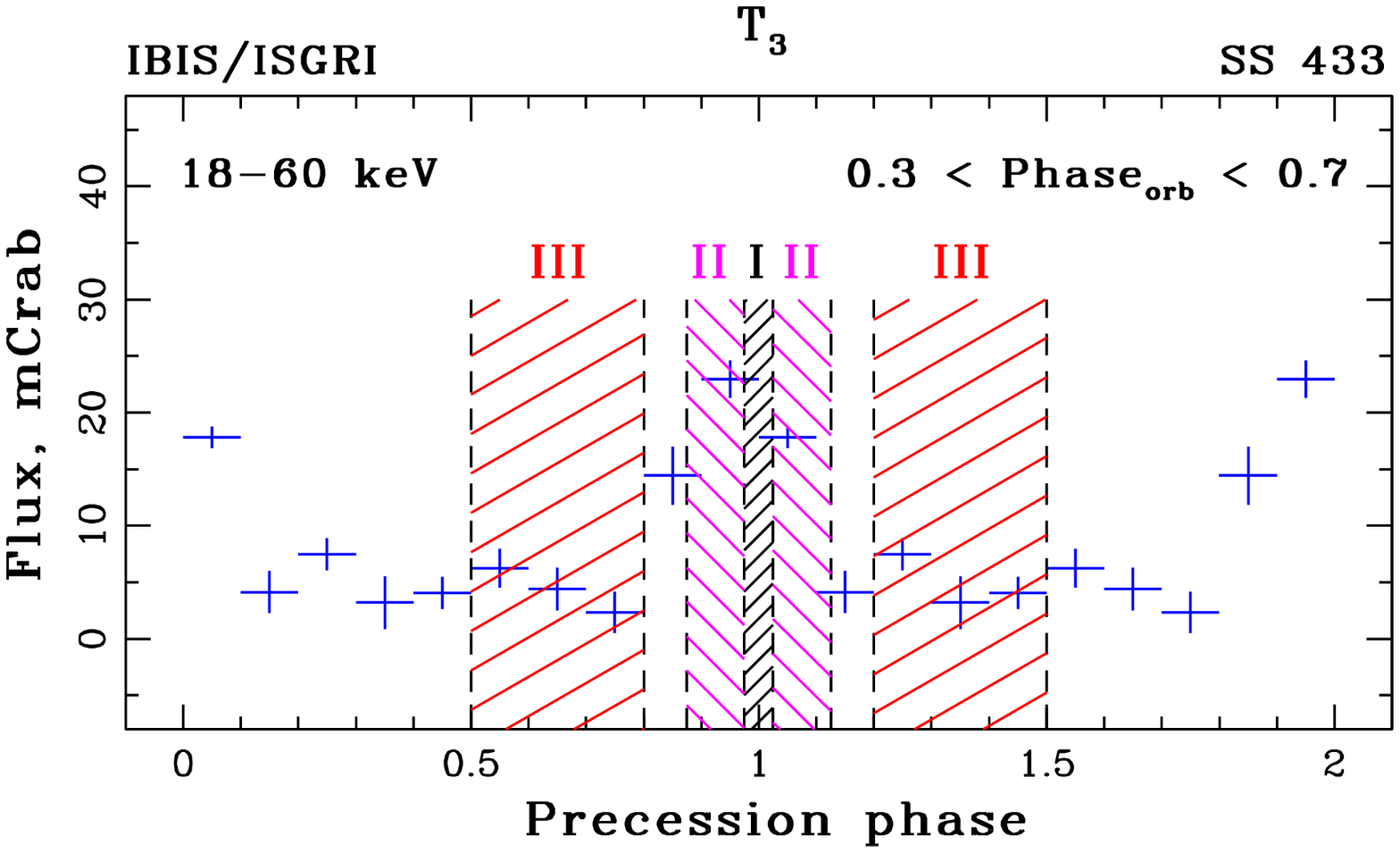}
\parbox[t]{0.47\textwidth}{
\caption{Average precessional light curve of SS433 from all available IBIS/ISGRI 18-60~keV data. Orbital eclipses are excluded.\label{f0}}
}
\hfill
\parbox[t]{0.47\textwidth}{
\caption{Precessional phase intervals chosen for spectral analysis of SS433.
\label{f1}}
}
\end{figure}

In this Section we focused on the precessional phase-resolved 
analysis of SS433.
To exclude the orbital modulation, 
we have used only observations at orbital phases from the range
$0.3<\phi <0.7$. After this filtering we still have $\sim2.8$~Ms
of data distributed across all precessional phases,
as shown in Fig. \ref{distr_tot_exp} (the right panel).  

The combined
AO1 -- AO5 data (including those publically available) 
enabled us to measure, for the first time, the precessional 
hard X-ray variability (Fig. \ref{f0},\ref{f1}), which 
turned out to be quite significant and stable over several 
precessional periods. The maximum to minimum flux ratio of the average precessional variability 
is around 5-7, which is higher than in softer X-ray bands, and suggests 
the hard X-ray emission originating closer to the base of the visible part of the jets.  
The eclipsing and precessional variabilities, combined with 
spectroscopic data simultaneously taken with INTEGRAL observations by the 6-m telescope SAO RAS, were taken into account when attempting to model the light curve of the source by \citep{Cher05}. 
However, the results 
proved to be inconclusive. First of all, the spectral resolution of the optical observations (around 3000) was insufficient to definitely measure the radial velocity curve of the optical companion.  
The modeling of X-ray eclipse and precessional variability allowed fairly broad range of parameters, including the mass ratios $q\sim 0.1-0.5$ (primarily due to the 
lack of reliable observations at the middle of 
the hard X-ray eclipse). An independent spectral analysis of archival  RXTE observations of SS433 \citep{Fil06} also suggested that it is hard to derive the binary mass ratio
based on X-ray eclipse observations only.

\subsection{Hard X-ray spectra}
\label{s_hardsp}

\begin{figure}
\includegraphics[width=0.9\textwidth]{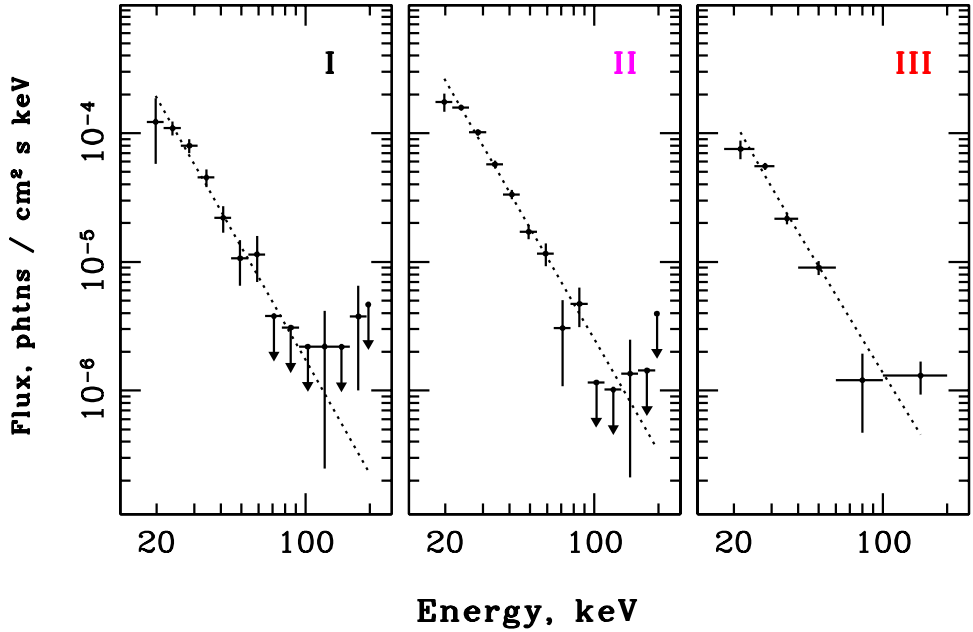} 
\caption{IBIS/ISGRI $10-200$~keV SS433 spectra at different precessional phases. 
From left to right: spectra averaged 
inside phase intervals I, II, III shown in Fig. \ref{f1}, respectively. All spectra can be fitted by a power-law model with the photon spectral index $-2.8$ (the dotted line).}
\label{Xspectra}
\end{figure}

Our previous studies \citep{Cher03,Cher05,Cher06} suggested that SS433 is 
a superaccreting galactic microquasar with hot corona above the accretion 
disk which scatters hard X-ray radiation from the central parts of the disk.
Out of the eclipse, the source has been reliably detected 
by IBIS/ISGRI telescope up to 100 keV, 
with the X-ray continuum fitted by a featureless power-law $dN/dt/dA/dE\sim 
E^{-2.8}$~ph/s/cm$^2$/keV \citep{Cher03, Cher05, Cher06}, see Fig. \ref{Xspectra}.

INTEGRAL observations (2003-2007) were separated into 
three segments in the precessional phase (Fig. \ref{f1}) (limited by photon statistics). 
The resulted spectra are shown in Fig. \ref{Xspectra}, with no clear difference in the power-law fit
in the 20-100 keV energy band. It is recognized from RXTE observations 
that at softer X-ray band 
the spectrum is very variable \citep{Fil06}, but errors in the first point 18-22 keV of
our fit which can be due to this variability have no effect on the value of the power-law fit.
One hard X-ray spectral fit at different precessional phases of the accretion disk
suggests the presence of 
a fairly broad hard X-ray emission region.  To be independent of the 
disk opening angle varying with the precessional phase, the size 
of this region must be 
comparable to that of the accretion disk ($10^{11}-10^{12}$ cm).

The X-ray continuum spectrum of SS433 in the broad range 3-100 keV 
can be fitted by a two-component model including thermal X-ray emission from 
the cooling jet and thermal comptonization spectrum from the isothermal corona
\cite{Krivosheev08}.
For both uneclipsed and partially eclipsed by the accretion disk corona
the model gives the best spectral fits
for $T_{cor}\simeq 20$~keV with Thomson optical depth $\tau_T\simeq 0.2$,
the mass outflow rate in the jet $\dot M_j\simeq 3\times 10^{19}$~g/s
corresponding to a jet kinetic power of $\sim 10^{39}$~erg/s.
These parameters suggest an electron density of $\sim 5\times 10^{12}$~cm$^{-3}$
at distances $\sim 5\times 10^{11}$~cm. 
Such a density is typical in the wind 
outflowing with velocity $v\sim 3000$~km/s from a supercritical 
accretion disk with $\dot M\sim 10^{-4}$~M$_\odot$/yr at distances $\sim 10^{11}-10^{12}$~cm from 
the center, where the Compton-thick photosphere is formed \citep{Fab04, Revnivtsev04}.

Note that despite several primary eclipses have already been observed by INTEGRAL, the
measured hard X-ray flux $\sim 5-20$~mCrab from the source is still insufficient 
for phase-resolved spectral analysis of an individual eclipse. To this aim, additional
observations of SS433 by INTEGRAL are planned.

\subsection{The primary eclipse and its peculiarity}
\label{s_primecl}

The dedicated INTEGRAL observations of the primary eclipses of SS433 
in 2003, 2007 and 2008 were obtained near the $T3$  
precessional phase when the accretion disk is maximally open. The data include:

 I. INTEGRAL orbits 67-70 (May 2003), precessional phase $\psi=0.001-0.060$;

 II. INTEGRAL orbits 555-556 (May 2007), precessional phase $\psi=0.980-0.014$;

 III. INTEGRAL orbits 608-609 (October 2007), precessional phase $\psi=0.956-0.990$;

IV. INTEGRAL orbits 612-613 (October 2007), precessional phase $\psi=0.030-0.064$. 

V. INTEGRAL orbits 722-723 (September 2008), precessional phase
$\psi=0.057-0.091$.

\begin{figure}
\includegraphics[width=0.47\textwidth]{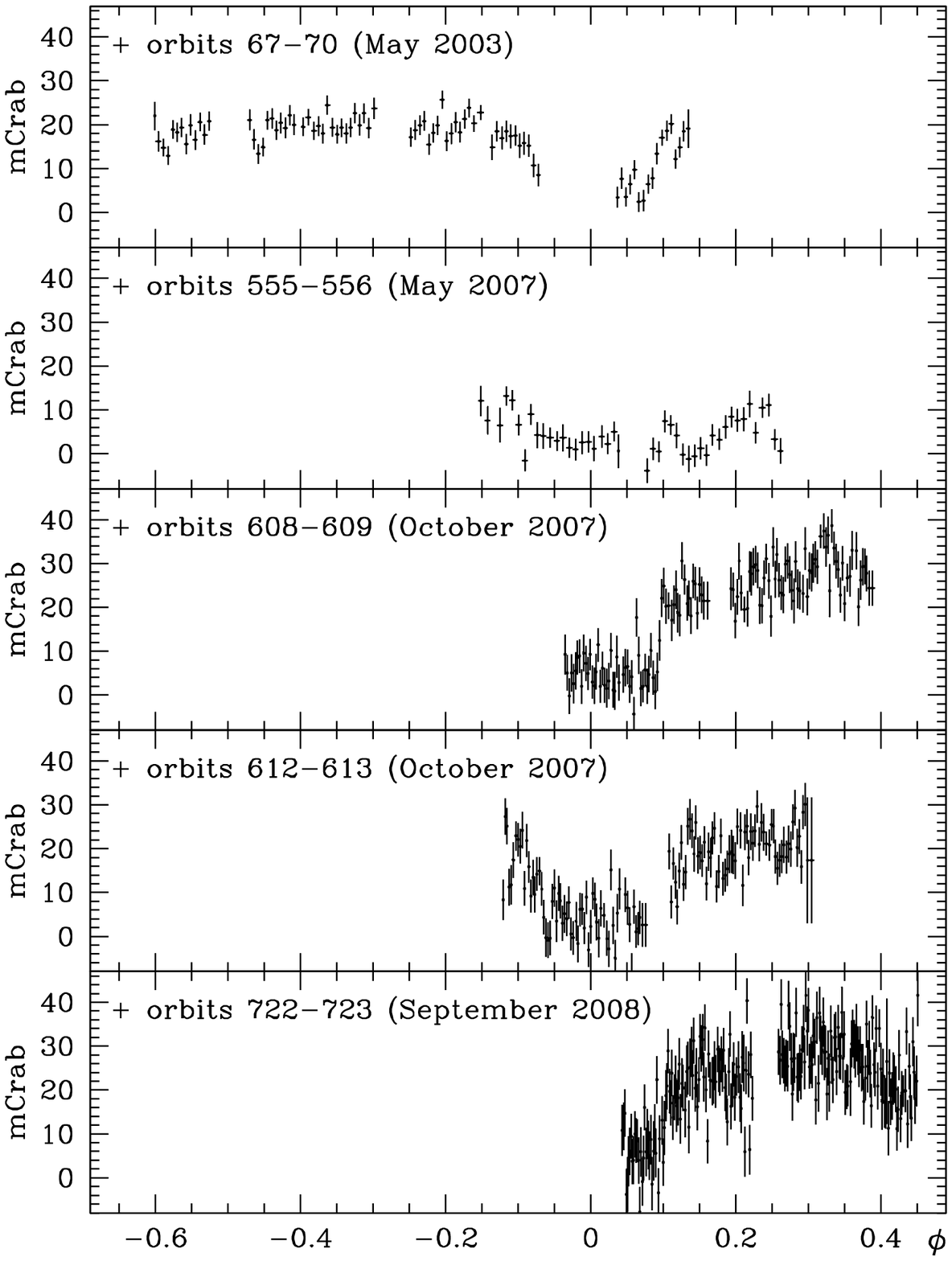}
\hfill
\includegraphics[width=0.5\textwidth]{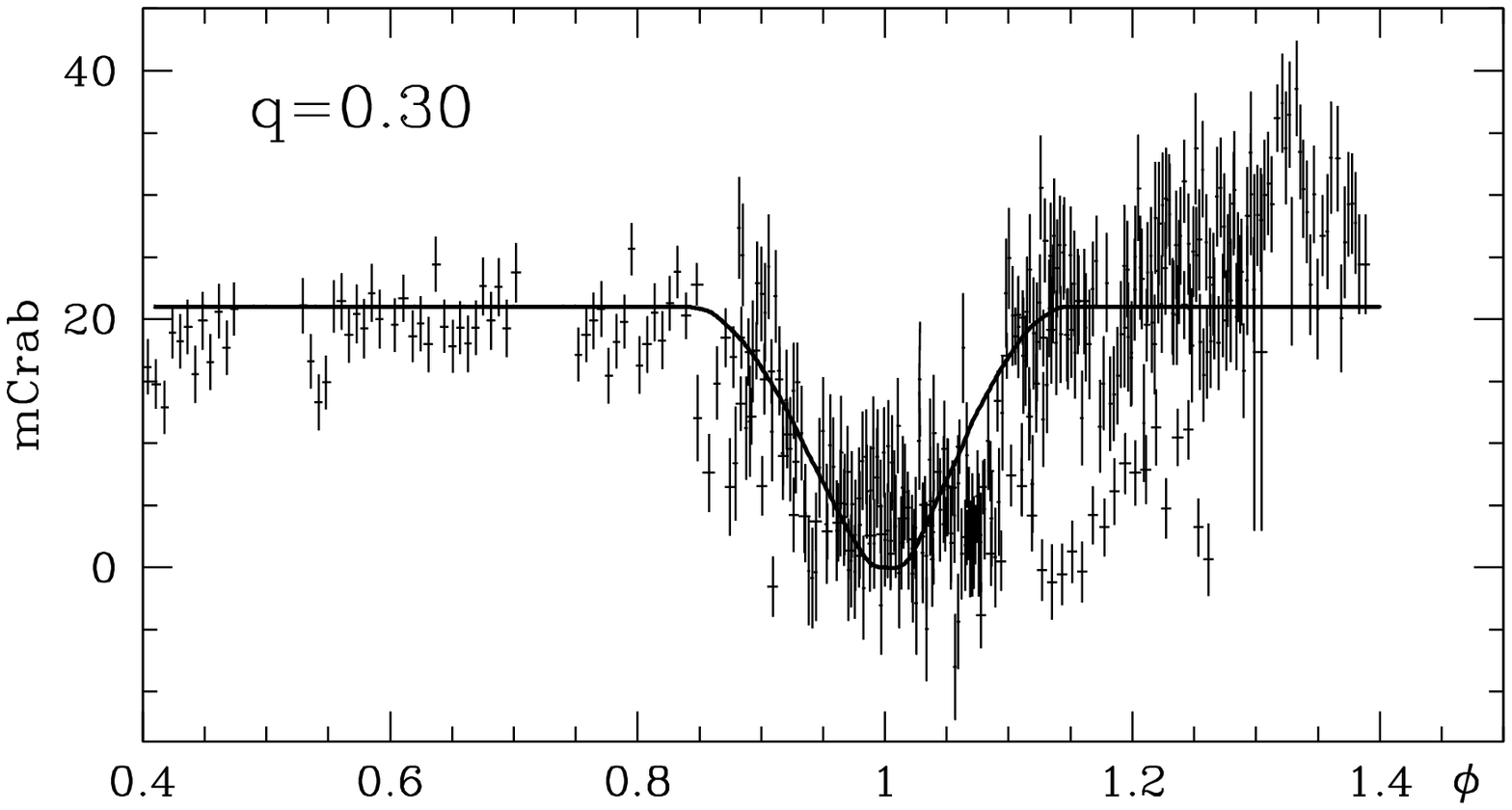}
\parbox[t]{0.47\textwidth}{\caption{From top to bottom:  
primary eclipses of SS433 observed by INTEGRAL (IBIS/ISGRI data, 18-60 keV) during I, II, III, IV and V sets (see the text).\label{f_alleclipses}}}
\hfill
\parbox[t]{0.47\textwidth}{\caption
{Combined I-IV INTEGRAL primary eclipses of SS433. The solid
curve shows the model light curve for $q=0.3$ (see the text).}\label{eclipse}}
\end{figure}

The observed eclipses are shown in Fig \ref{f_alleclipses}. In our analysis
we have used only sets I, III, IV, V and excluded set II (the one showing 
the most suppressed egress out of the primary eclipse).  


SS433 is known to be a highly variable object at all wavelengths \citep{Cher88,Fab04,Revnivtsev04,Revnivtsev06}. This is not surprising for a supercritically
accreting massive binary, considering various instabilities that should be immanently present in the
mass outflow from the optical star and in the accretion disk. 
It might appear that the primary eclipse of the hard X-ray emission 
coming from the regions close to the 
central black hole by the optical star should be stable. However,
it is not the case for hard X-ray eclipse in SS433 (Fig. \ref{f_alleclipses}). As seen in this figure, the second eclipse observed in May 2007 (INTEGRAL orbits 555-556) has an unusually broad form: a shallow ingress to and broad egress from the eclipse, which may be  due to the low level of uneclipsed X-ray flux from the source $\sim 10$~mCrab. Note that this particular eclipse is 
very similar to the one observed by \textit{Ginga} \citep{Kawai89}. However, 
already at the next T3 moment in October 2007 (sets III and IV in Fig. \ref{f_alleclipses}) the uneclipsed X-ray flux was at a level of 20 mCrab and the egress from eclipse restored its more familiar form (like in set I). 
This effect is particularly clearly visible on the combined X-ray eclipse light curve shown in Fig. \ref{eclipse}. The reason for such a strong variability of the hard X-ray egress may be
related to powerful inhomogeneous gaseous streams feeding the disk and the wind from 
the supercritical accretion disk. Additional X-ray absorption 
can be produced in the collision region of the winds from the optical star and the supercritical accretion disk \citep{Cher95}.

The slow dense wind from the accretion disk in SS433 
should form a Compton-thick 
photosphere and funnels around relativistic jets \citep{Fab04}. The height of the 
funnel $\sim 10^{12}$~cm, comparable to the accretion disk size, 
can be inferred from the observed $\sim 80$-s lag of the optical emission with respect to the X-rays from the jet base found in the simultaneous optical/X-ray observations of the source
\citep{Revnivtsev04}. Thus it appears quite possible that the variable wind from the precessing supercritical accretion disk 
causes changes of the funnel structure, so even at the maximum opening disk
phase (T3) the shape of hard X-ray eclipse may appreciably vary depending 
on the accretion rate through the disk. In this connection it is worth
noticing the variable broad absorption-like feature that appears immediately after
the egress from the eclipse (at orbital phases $\phi\sim 0.15-0.2$, 
see Fig. \ref{f_alleclipses}). It would be very interesting to know whether this
feature is due to the true absorption of X-ray emission or just reflects the strong variability of the proper X-ray flux. 

We conclude that the hard X-ray eclipse in SS433 is likely to be formed by 
both geometrical screening of the broad X-ray emitting region by the opaque
star and complex gaseous stream and/or inhomogeneous wind from the optical star and the supercritical accretion disk, as well as by the wind-wind collision region. This implies that when fitting the hard X-ray primary eclipse in SS433 by geometrical model, 
we should use the most stable part of the observed eclipses, i.e. 
the ingress to the eclipse and the \textit{upper
envelope} of the observed egress out of eclipse.  
In view of high variability, the analysis of one individual X-ray eclipse only 
can lead to erroneous determination of the binary system parameters. Moreover, for their reliable determination a joint modeling of both orbital and precessional variability of hard X-ray flux is needed.  

\section{Analysis of hard X-ray light curves of SS433}

\subsection{Geometrical model of SS433}
\label{s_geommod}

To analyze hard X-ray eclipses of SS433 we used a geometrical model 
developed earlier for the interpretation of the {\it Ginga} data \citep{Ant92}
and the {\it INTEGRAL} light curve \citep{Cher05}.
We consider a close binary system consisting of an (opaque) "normal"  star
limited by the Roche equipotential surface and a compact object
surrounded by an optically and geometrically thick "accretion disk".
Relativistic jets are directed perpendicularly to the disk plane.
The "accretion disk" includes the disk itself and an extended photosphere
formed by the outflowing wind. The orbit is circular, the axial rotation
of the normal star is assumed to be synchronized with the orbital revolution. 

\begin{figure}
\centering
\includegraphics[width=0.5\textwidth]{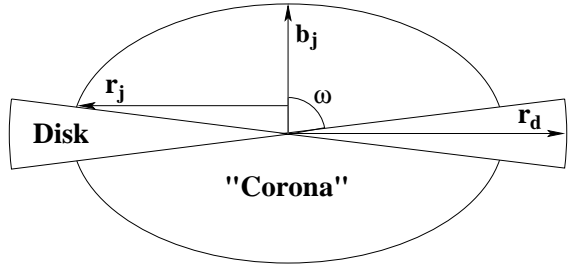}
\caption{Geometrical model of the accretion disk and its "corona".}
\label{f_geommodel}
\end{figure}

The disk and "jets" are precessing in space and change the orientation
relative to the normal (donor) star. 
The disk is inclined with respect to the orbital plane by the angle $\theta$.
A cone-like funnel is located inside the disk and is characterized by the
half-opening angle $\omega$, thus the opaque disk body (see Fig. \ref{f_geommodel}) is
described by the radius $r_d$ and the angle $\omega$.
The central object is surrounded by a transparent quasi-isothermal 
homogeneously emitting spheroid with a visible radius $r_j$ and height $b_j$
which could be interpreted as a "corona" or a "thick jet" (without any 
relativistic bulk motion). Here $r_j$, $b_j$ and $r_d$ are dimensionless values
expressed in units of the binary separation $a$.
The radius of the normal star is determined by the relative Roche lobe size,
i.e. by the mass ratio $q=m_x/m_v$ ($m_x$ is the mass of the
compact object).

Only the "corona" is assumed to emit in the hard X-ray band, while the star
and disk eclipse it in the course of the 
orbital and precessional motion. During
precession the inclination of the disk with respect to the observer changes,
causing different visibility conditions for the "corona".
The "corona" is partially screened by the cone disk edge, which can
explain qualitatively the change of the uneclipsed X-ray flux with
precession phase. 
Observations of the 
precessional variability can thus be used to obtain a "vertical" scan of the
emitting structure, restricting the parameters $b_j$  and $\omega$.
The orbital (eclipse) variability observations scan the emitting
structure "horizontally", restricting possible values of $r_d$, $\omega$, $q$ and
$r_j$. The joint analysis of the precessional and eclipse variability thus enables us
to reconstruct the spatial structure of the accretion disk
central region, where the hard X-rays are produced, 
and to estimate the binary mass ratio $q$.

The position of the components of the system relative to the 
observer is determined by the binary orbit inclination angle $i=78.8^o$, 
the disk inclination angle to the orbital plane $\theta=20.3^o$ \citep{Margon84, Davydov08}, and 
the precessional phase $\psi$. 

\subsection{Results of light curves analysis}
\label{s_analysis}

Our model for the orbital and precessional variability of SS433 has
the following free parameters: the binary mass ratio $q=m_x/m_v$ determining the relative 
size of the normal star, 
the disk parameters $r_d, \omega$ determining the form of the disk, 
the thick "jet" or "corona" parameters $r_j, b_j$ determining the form of the hot corona. Note that the X-ray emitting region is qualitatively different for 
a thin narrow "jet" with $r_j\ll b_j\sim a$ and a thin short "jet" with $r_j\ll  b_j\ll a$. If $r_j>b_j$, it is more appropriate to term it as a corona (or
a thick "jet"). For each value of $q$ from the range $0.05-1.0$, we found other parameters best-fitting simultaneously the orbital 
and precessional light curves. We remind that the precessional light curve
was constructed using maximum off-eclipse X-ray flux from the system. 
The precessional variability amplitude was
assumed to be $A_{pr}\simeq 5-7$. The $\chi^2$ criterion was used to evaluate
the goodness of fits. 

To examine as fully as possible the binary system parameters, we have used 
different variants of modeling of the observed orbital hard X-ray variability. These included:

1) the analysis of light curves consisting of individual points;

2) the analysis of average light curves;

3) the analysis of individual observational sets (I, III, and IV-V);

4) the analysis of the combined data from sets I, III and IV-V;

5) the analysis of the entire eclipse, including the eclipse ingress and (the upper envelope of) egress;

6) the analysis of the ingress to the eclipse only (orbital phases $\psi_{orb}=0.8-1.0$).

All methods give similar results. 
The inferred binary system parameters  
are found to be sensitive mostly to the form of ingress and the duration of the
primary eclipse. The results of the analysis can be summarized as follows.


1) The analysis of individual and average light curves yields similar results
(i.e. the use of the average light curve does not add errors
to the obtained parameters). This might seem obvious, but we 
directly checked it. 

2) The analysis of individual data sets I, III, IV, V 
and the combined data yields similar results. 

Taking into account the above conclusions, below we shall show the model
light curve for the combined data sets I, III, IV and V.

3) When searching for the best-fit model parameters at different $q$, 
we found that solutions obtained for the entire primary eclipse
(the ingress and the upper envelope of egress) and for 
the eclipse ingress only are not very different. Taking into account highly
variable character of the eclipse egress noted above, 
below we shall discuss solutions and model parameters 
obtained from the analysis of the \textit{eclipse ingress part only}. 
In the combined primary eclipse (Fig. \ref{eclipse}) the ingress part 
is mostly determined by sets I and IV. 

4) In our previous analysis of the primary orbital eclipse observed by INTEGRAL \citep{Cher05} we found that for all models the best-fit values 
(corresponding to the minimum $\chi^2$) are obtained for the maximum possible radius of the disk $r_d$ and the maximum possible value of the angle $\omega$. This conclusion is confirmed here using new data. We also confirm that parameters $b_j$ and $\omega$ are correlated. 

\begin{figure}
\includegraphics[width=0.5\textwidth]{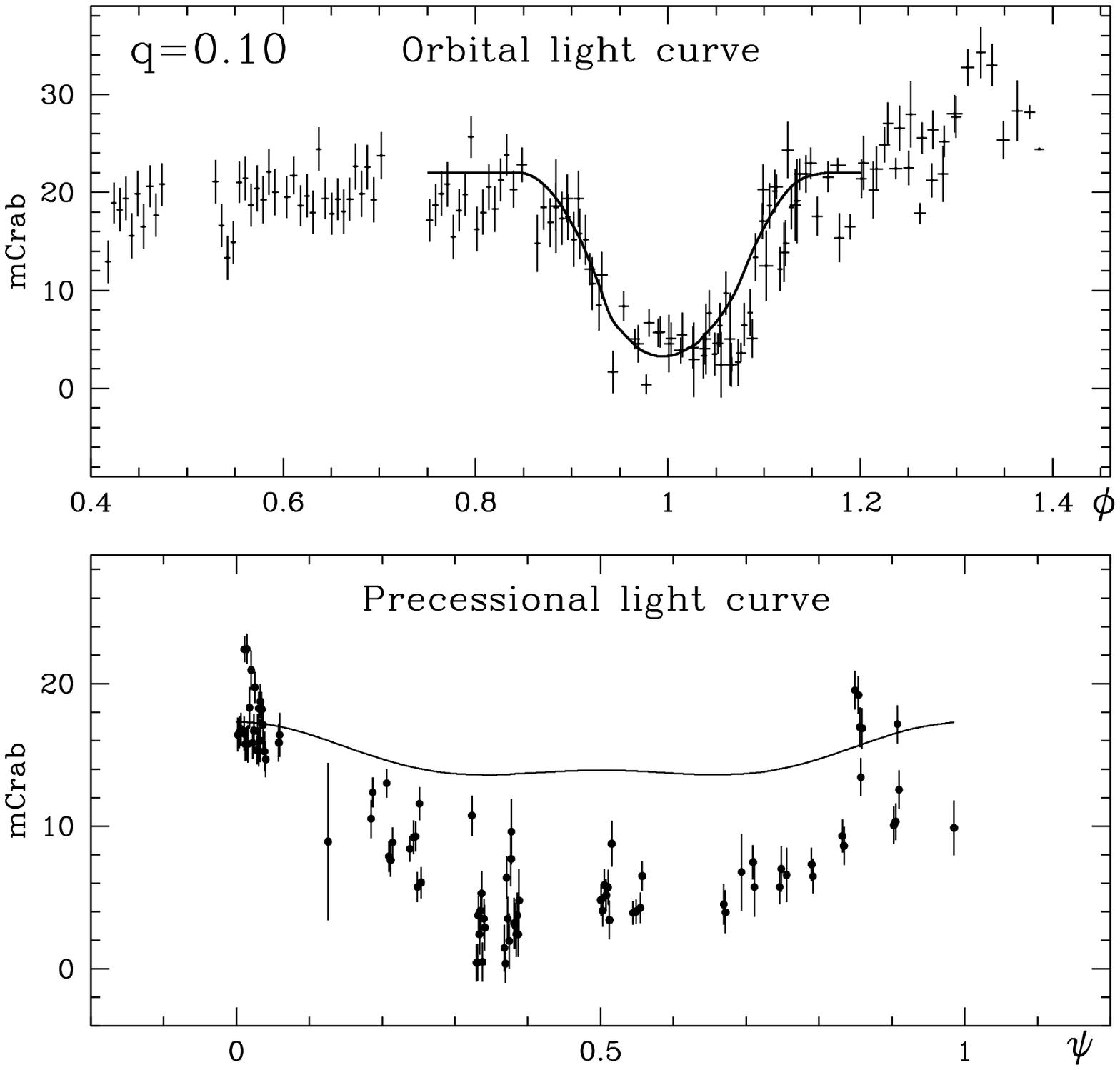} 
\hfill
\includegraphics[width=0.5\textwidth]{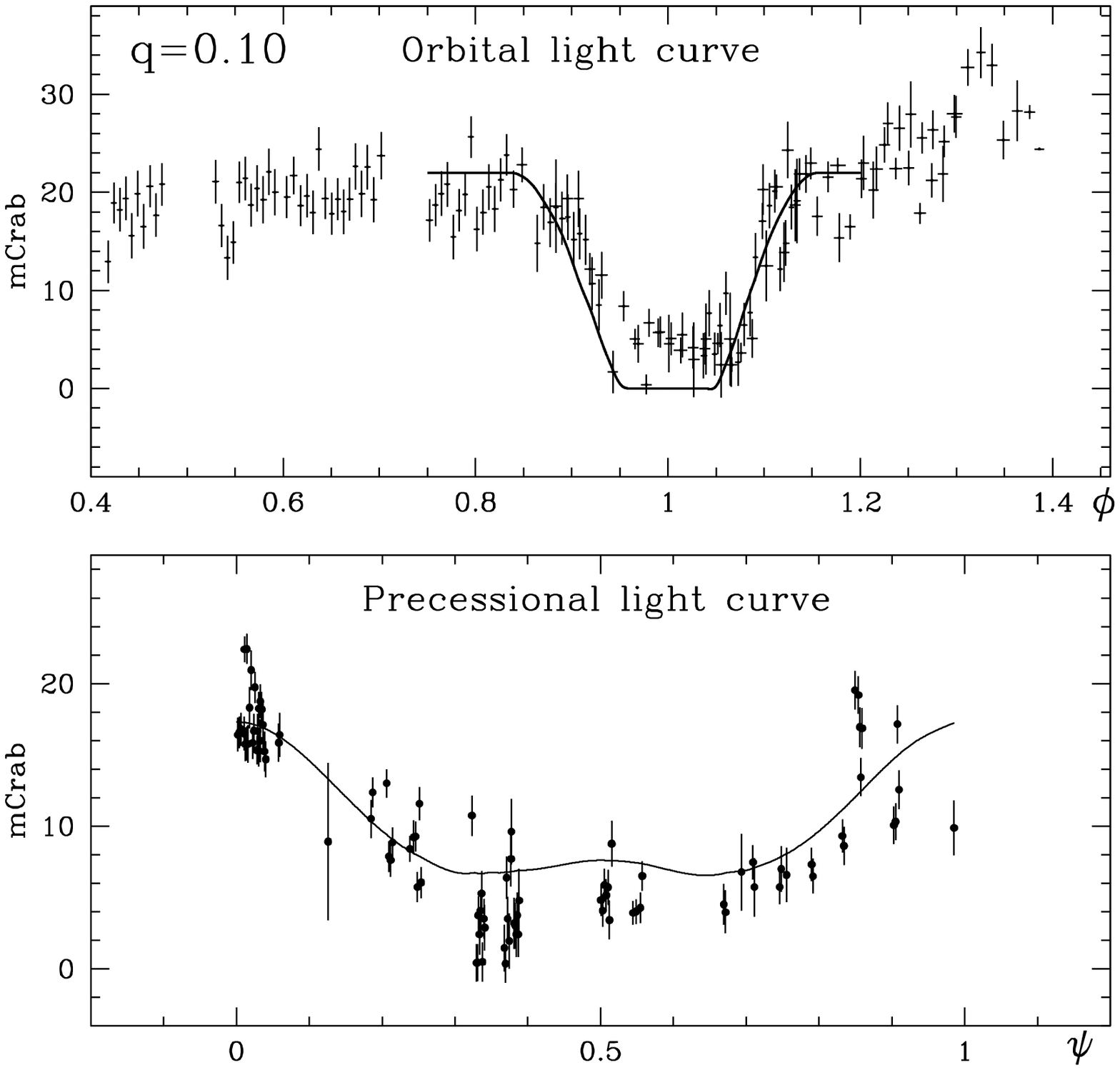}
\parbox[t]{0.47\textwidth}{\caption{Best-fit to the orbital
(top) and precessional (bottom) light curve 
for small binary mass ratio $q=0.1$ by \textit{long} X-ray jet.}\label{f_longjet}}
\hfill
\parbox[t]{0.47\textwidth}{\caption{
Best-fit to the orbital
(top) and precessional (bottom) light curve 
for small binary mass ratio $q=0.1$ by \textit{short} X-ray jet.} 
\label{f_shortjet}}
\end{figure}

5) The minimum deviation of the best-fit model eclipsing light curve from observed points is reached for the small mass ratio $q\sim 0.1$ and a long X-ray emitting "jet" ($b_j>0.5$) with the base radius varying in the wide range ($r_j=0.05-0.25$). This means that the "jet" can be long but either thin or thick. However, the main objection to model of the long "jet" comes from its inability to describe
the observed precessional variability as shown in Fig. \ref{f_longjet}. 
We conclude that at small mass ratio $q\le 0.25$ our
model does not simultaneously fit both orbital and precessional variability of SS433 observed by INTEGRAL. The precessional light curve could be fitted at the small mass ratio ($q=0.1$) by a short X-ray emitting "jet", but then the flux at the center of the primary eclipse would be zero 
(Fig. \ref{f_shortjet}), again  
in contradiction with observations -- in all cases, INTEGRAL detects a non-zero X-ray flux of $\sim 3$~mCrab at the middle of the hard X-ray eclipse (Fig. \ref{f_alleclipses}).

\begin{figure}
\includegraphics[width=0.5\textwidth]{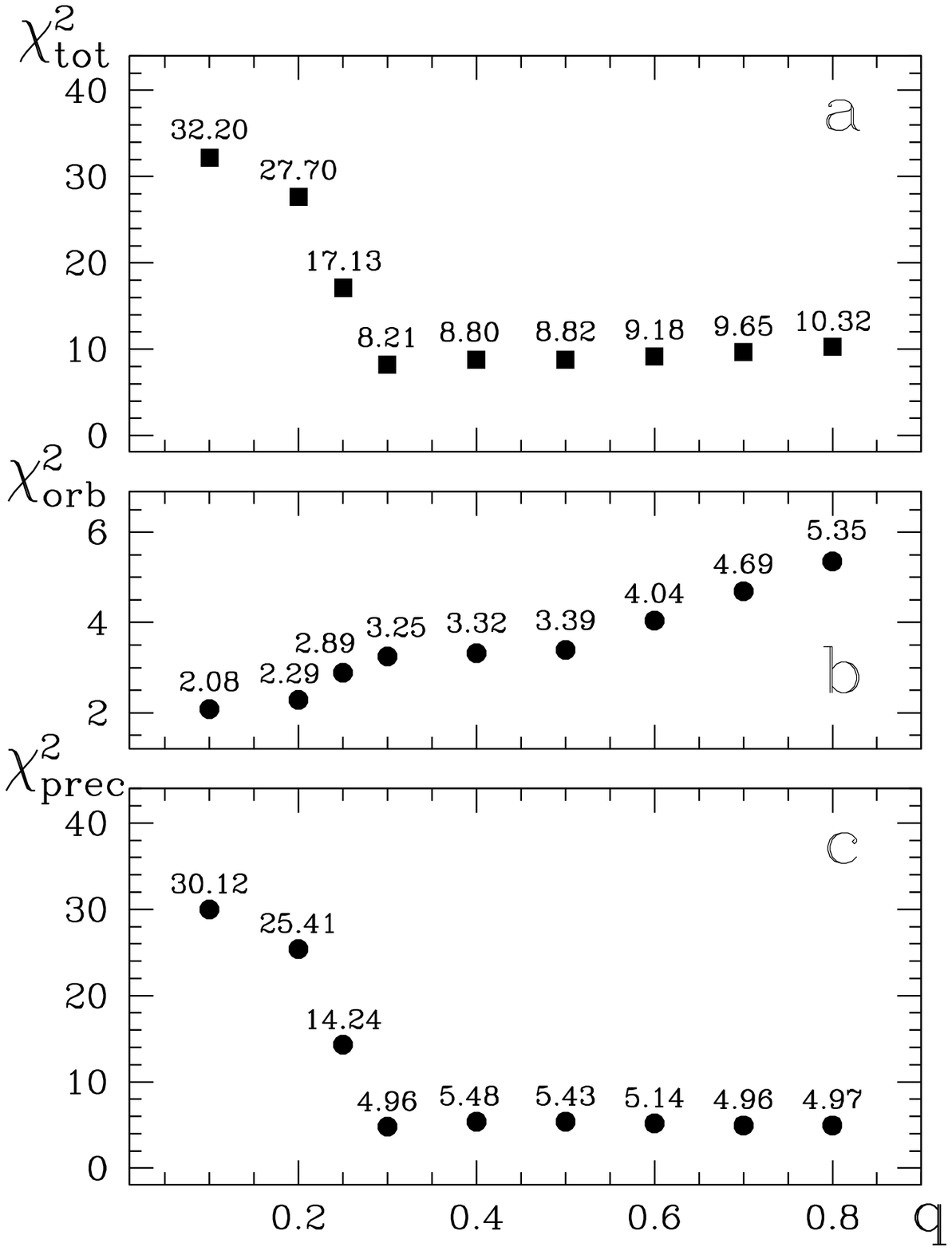}
\hfill
\includegraphics[width=0.5\textwidth]{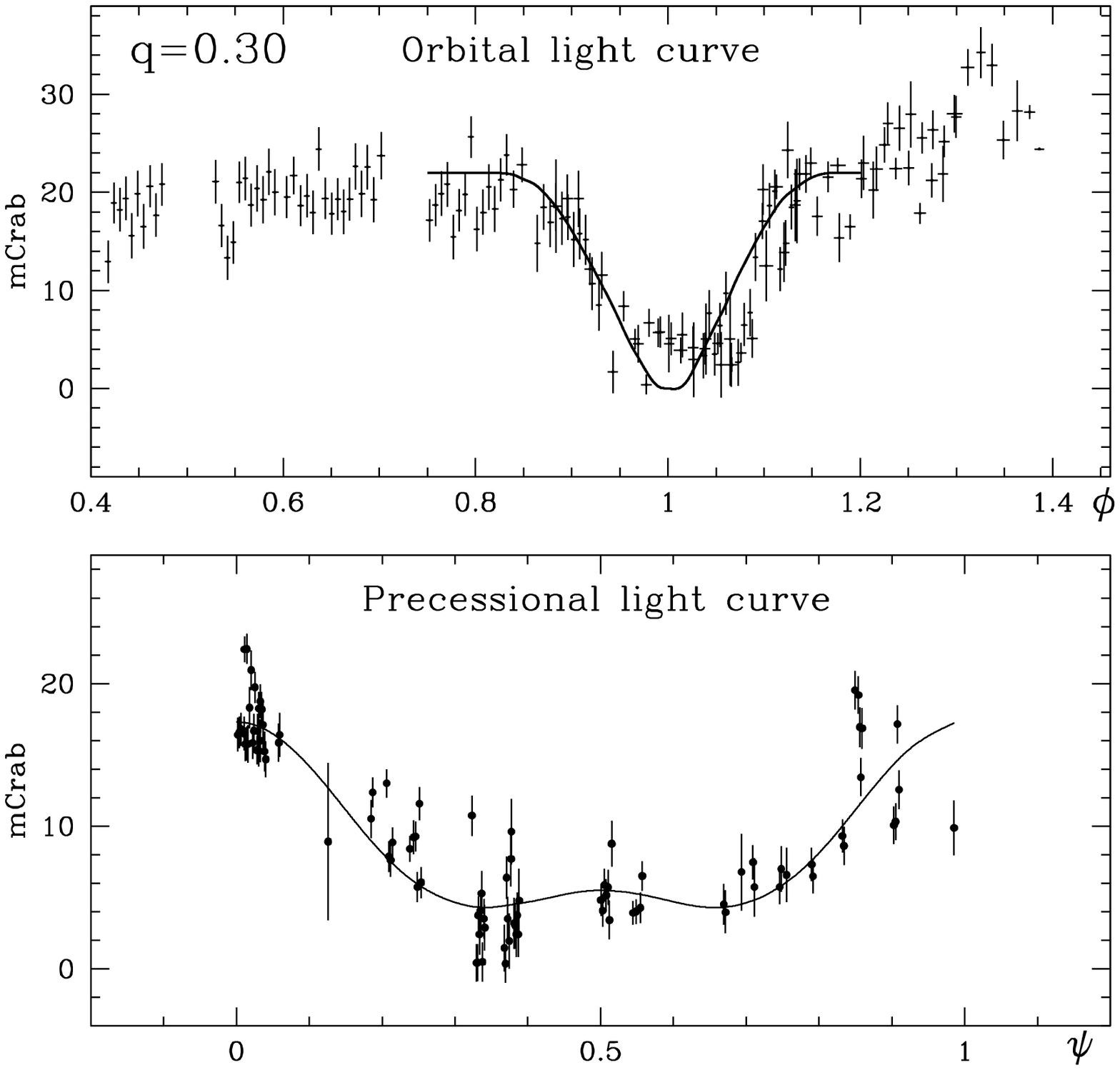} 
\parbox[t]{0.47\textwidth}{\caption{
Panel a: the combined reduced 
$\chi^2_{tot}=\chi^2_{orb}+\chi^2_{prec}$ (the sum of the $\chi^2$ for the best-fit orbital light curve 
and the corresponding precessional light curve) for different $q$. The reduced values 
of orbital eclipse 
$\chi^2_{orb}$ and precessional $\chi^2_{prec}$ variability for different $q$ are shown in panel (b) and (c), respectively. Figures mark the value of the corresponding $\chi^2$. }\label{chi2}}
\hfill
\parbox[t]{0.47\textwidth}{\caption{The best-fit orbital (top) and
precessional hard X-ray light curve of SS433 for binary mass ratio $q=0.3$.}\label{q03}}
\end{figure}

\begin{figure}
\includegraphics[width=0.5\textwidth]{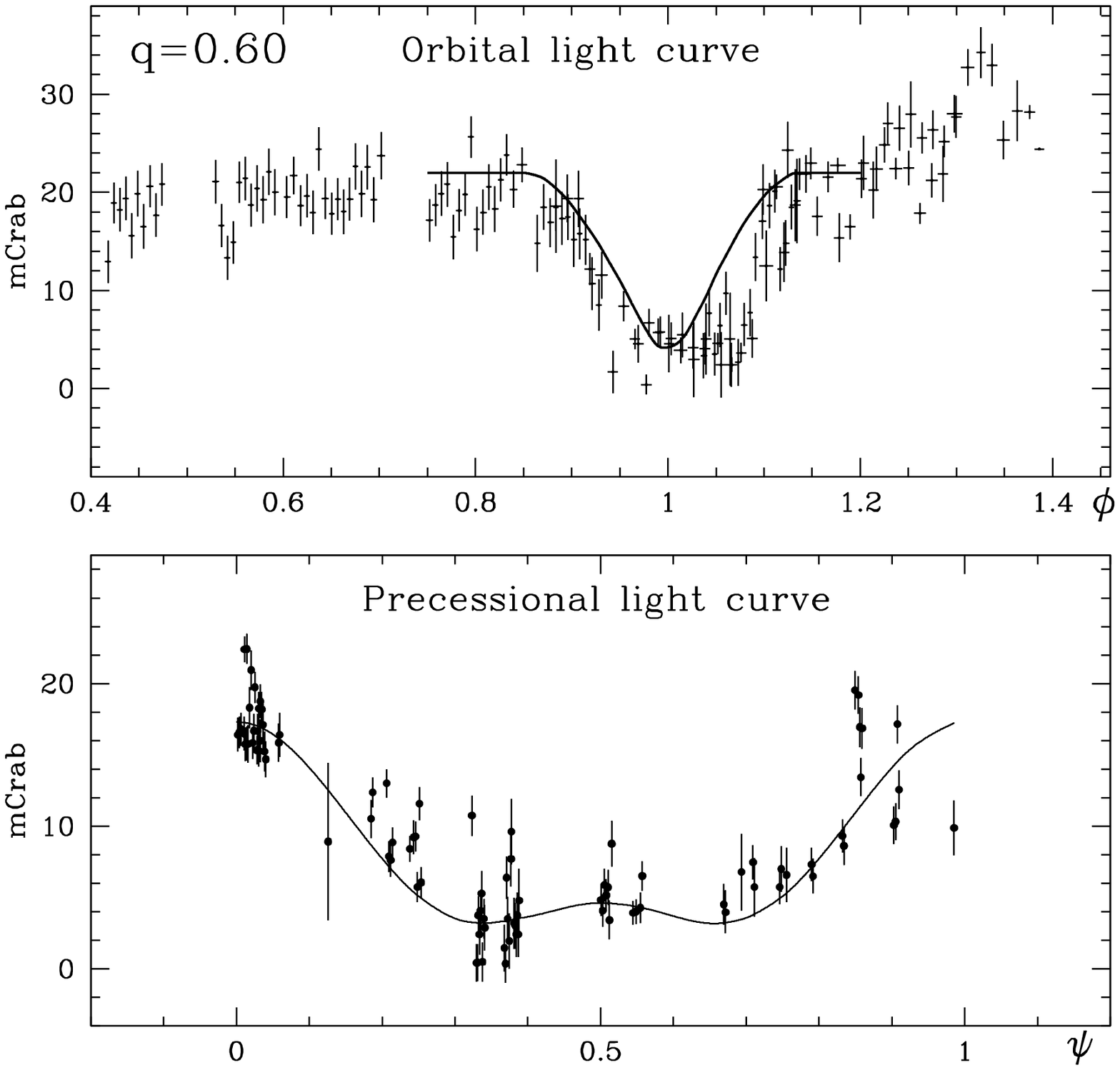}
\hfill
\includegraphics[width=0.5\textwidth]{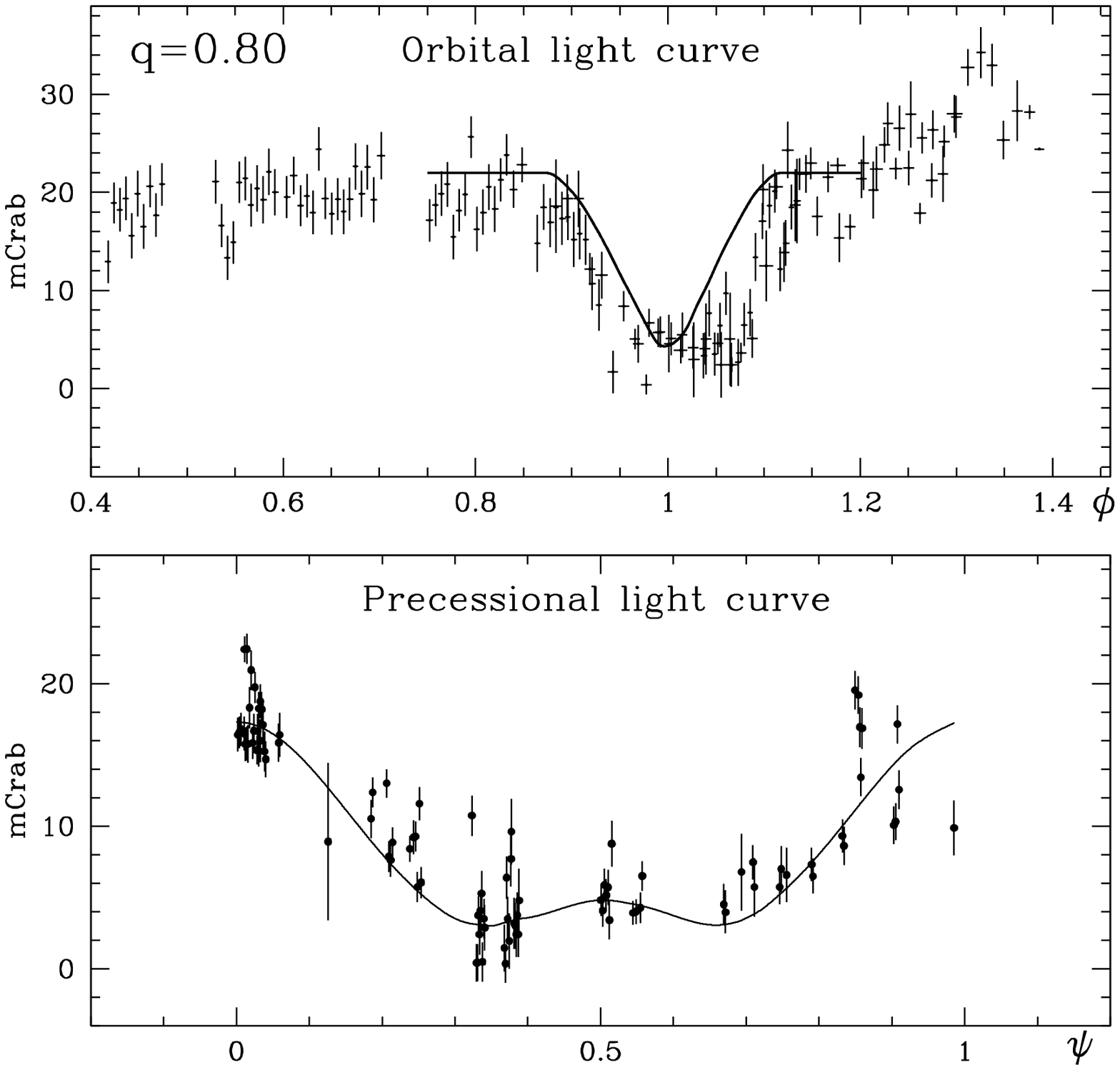}
\parbox[t]{0.47\textwidth}{\caption{The same as in Fig. \ref{q03} for $q=0.6$.
\label{q06}}}  
\hfill
\parbox[t]{0.47\textwidth}
{\caption{The same as in Fig. \ref{q03} for $q=0.8$.}\label{q08}}
\end{figure}

6) For $q=0.3-0.5$ our model provides 
almost similarly good fits to the orbital and precessional variability of SS433. To find the best-fit value of $q$ we 
calculate the sum of the reduced $\chi^2$ for the orbital and precessional
light curves obtained for the same model parameters:
$\chi^2_{tot} = \chi^2_{orb} + \chi^2_{prec}$ (Fig. \ref{chi2}, panel (a)). 
The residuals are minimized relative to four parameters: $r_d, \omega, r_j, b_j$. The value
$\chi^2_{orb}$  (Fig. \ref{chi2}, panel (b)) 
corresponds to the best-fit 
parameters of the orbital light curve for  different $q$.  The value $\chi^2_{prec}$ is
calculated for the corresponding precessional light curve (Fig. \ref{chi2}, panel (c)). 
Fig. \ref{chi2}, panel (b), shows that 
$\chi^2_{orb}$ deviations for the orbital eclipse increase with $q$; there is 
a quasi-plateau at $q\sim 0.3-0.5$, then the deviations rapidly increase due to a poor fit
to the orbital eclipse (especially the eclipse width, see Fig. \ref{q06}). 
Fig. \ref{chi2}, panel (c),  
demonstrates that for $q < 0.25$ fits to the precessional light curve sharply worsen. 

As follows from Fig. \ref{chi2}, the values of the reduced $\chi^2_{tot}$ and 
$\chi^2_{orb}$ (minimized relative to four model parameters $r_d, \omega, r_j, b_j$) 
are much higher 
than unity. This means that our model is oversimplified 
and is not fully adequate to the observational data. The data demonstrate a significant dispersion due to both physical variability of hard X-ray flux and systematic effects of absorption by gaseous streams and in the wind-wind collision region, which we do not take into account in our model. 
So we cannot rigorously estimate the confidence region  
of the sought parameter $q$. When the reduced $\chi^2>1$, 
it is generally accepted to give only the optimal value of the 
sought parameter, without quoting the confidence range since the latter 
can not be statistically justified. 
The absolute  minimum of the residuals $\chi^2_{tot}\simeq 8.21$ is reached at 
$q\simeq 0.3$, so we accept it as  the optimal value.
Nevertheless, as the reduced residuals $\chi^2_{orb}$ and  
$\chi^2_{prec}$  strongly change with $q$ (Fig. \ref{chi2}), 
it appears meaningful to consider the mass ratio $q$ from the range $\sim 0.25-0.5$, where 
the orbital $\chi^2_{orb}$ shows a plateau. The 
lower limit of the mass ratio interval is determined by the abrupt increase in 
the precessional $\chi^2_{prec}$ residuals for $q<0.25$, the upper limit is determined by 
(less abrupt) increase in $\chi^2_{orb}$ residuals for $q>0.5$.
%
%
. 

Therefore, our simple geometrical model of SS433 suggests an acceptable 
range of binary mass ratio of $0.25\le q\le 0.5$. The absolute minimum of the total reduced residuals $\chi^2_{tot}\simeq 8.21$
at $q\simeq 0.3$ agrees well with the mass ratio estimate $q\simeq 0.35$ obtained by \cite{hillwig08}. We stress that 
the latter estimate is obtained from optical spectroscopy of SS433 and is in agreement with the estimate $q\simeq 0.4$ derived by \cite{AntCher87} from 
analysis of optical eclipse light curves at different precessional phases. Our new binary mass ratio estimate $q\simeq 0.3$ is obtained from independent observations in hard X-rays. 
Such a good convergence of the mass ratio estimates found from different 
and independent observational data strengthens the reliability of the found 
binary mass ratio value 
$q\simeq 0.3$.

The best-fit model parameters for $q\simeq 0.3$ (see Fig. \ref{q03})
correspond to an extended oblate corona with width comparable to the size of the accretion disk ($r_j\sim r_d$) and small vertical height ($b_j\sim 0.15-0.20$). In the framework 
of our model, such a 
geometry of the hard X-ray emitting region is caused by the 
wide primary eclipse minimum and large amplitude of the precessional variability. 
The precessional light curve in hard X-rays is shaped by the outer parts of the disk eclipsing the high-temperature corona that scatters thermal X-rays from jets and possibly
the innermost accretion disk \citep{Krivosheev08}. 

Fig. \ref{q03} 
shows the orbital and precessional 
light curves for the optimal binary mass ratio $q=0.3$. 
For comparison, best-fit model light curves for mass ratio
$q=0.6$ and $q=0.8$ are presented in Fig. \ref{q06} and Fig. \ref{q08}, respectively.
Figs. \ref{f_longjet}, \ref{f_shortjet}, \ref{q03}, \ref{q06} and \ref{q08} 
show
the change in the primary eclipse minimum width for different binary mass ratios.
Clearly, the model best-fit to the 
form and width of the primary eclipse at $q=0.6$ and $q=0.8$ is worse  
than that for $q=0.3$, as Fig. \ref{chi2} (the middle panel) formally shows.

\begin{figure}
\includegraphics[width=0.47\textwidth]{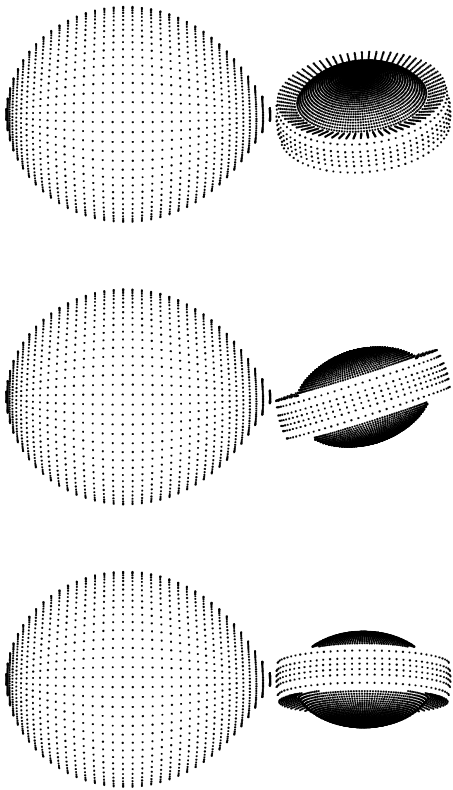}
\caption{Model view of SS433 for binary mass ratio $q=0.3$ at different 
disk precession angles (15, 118, 180 degrees from top to bottom, respectively).}
\label{f_q03_model}
\end{figure}

In Fig. \ref{f_q03_model} we 
illustrate the view of our optimal geometrical model of SS433 
with $q=0.3$ as seen at different precession angles. 
In this figure thin relativistic jets (emitting mostly in soft X-rays) are not shown. 

\section{Discussion}
\label{s_disc}

Our analysis confirms that SS433 is a superaccreting microquasar with black hole. The obtained high value of the best-fit binary mass ratio in SS433 $q\simeq 0.3$ allows us to explain a substantial amplitude ( $\sim 0^m.5$) of the optical variability at the minimum of the primary eclipse with precessional phase. The orbital inclination of SS433 is known from independent  analysis of moving emission lines ($i\simeq 78^\circ.8$, \cite{MargonAnderson89}), so at low mass ratios $q <0.25$ 
the small relative radius of the Roche lobe of the compact object should have caused the total eclipse of the bright precessing accretion disc around the compact object by the optical A7I-star, implying a constant optical flux at the center of the eclipse at different precessional phases. This contradicts observations (see e.g. \cite{Goransk98}). At $q=0.3\div 0.5$ the size of the Roche lobe of the compact object is relatively large to cause only partial primary optical eclipse of the accretion disc by the A7I-star at all precessional phases, which leads to a noticeable precessional variability of the minimum brightness at the middle of the primary eclipse. 

A small value of the binary mass ratio was found by \cite{Kawai89}
from the analysis of soft ($1\div 10$~keV) X-ray eclipses of SS433. The modeling of broad-band X-ray spectrum of SS433 shows that in this energy range thermal emission of relativistic jets with temperature decreasing along the jet dominates \citep{Fil06, Krivosheev08}. The observed broad width of the soft X-ray eclipse was interpreted by \cite{Kawai89}
in terms of purely geometrical model of the eclipse of jets by the optical star with sharp limb, yielding a small value of the binary mass ratio $q=0.15$. The observed independence of the soft X-ray eclipse width on energy has been the main argument justifying this model and hence the inferred low value of the mass ratio. However, this argument holds only if the temperature  along jets does not change with distance from the compact object. In fact it decreases along jets (unless some additional heating mechanism is assumed), so at energies $\sim 10$~keV central parts of the jets are eclipsed. Consequently, the duration of purely geometrical eclipses must be longer at 10 keV than at $\sim 1$~keV where cooler periphery of the jets is eclipsed \citep{Fil06}.  The independence of the X-ray eclipse duration on energy in the 1-10~keV range can be due to the compensation of the effect of decreasing temperature along jets and the increase of soft X-ray absorption in the extended atmosphere of the optical star, which is clearly seen 
in the RXTE observations \citep{Fil06}. So the effective radius of the eclipsing star increases at lower energies. This explains the constant width of X-ray eclipse in the 1-10 keV energy range. Turning this argument around, from the independent width of the X-ray eclipse in this energy range we can conclude that the eclipsing A7I-star in SS433 has not a sharp limb, so that X-ray eclipses in SS433 are not purely geometrical and are suffered from extinction in the heavy stellar wind from the optical star. This lends additional support to our interpretation of X-ray eclipses in SS433 making use of only the upper envelope of variable eclipsing hard X-ray light curve. 

Our analysis of the primary hard X-ray eclipse of SS433 
at the precessional phase 
$\psi\simeq 0.05$ with account for precessional variability of
the off-eclipse flux suggests the optimal binary mass ratio $q=m_x/m_v\simeq 0.3$. 
From spectroscopic observations of SS433 carried out by \cite{hillwig08}  
a semi-amplitude of the radial velocity curve of the optical A7I-component 
was determined: $K_v=58.2\pm 3.1$~km/s. Combining it  
with semi-amplitude of the radial velocity curve of 
the compact component $K_x=168\pm18$~km/s 
\citep{Hillwig_ea04}
and the 
binary inclination angle $i=78^\circ.8$ \citep{MargonAnderson89,
Davydov08}, 
we obtain for the mass function of
the optical component of SS433 
\begin{equation}
f_v(m)=\frac{m_x\sin^3i}{(1+1/q)^2}\simeq 0.268 M_\odot\,
\end{equation}
and the spectroscopic mass ratio $q\simeq 0.35$.
This allows us to estimate 
masses of the compact object and the optical star for
$q\simeq 0.3$:
\begin{equation}
m_x\simeq 5.3 M_\odot, \qquad m_v\simeq 17.7 M_\odot\,.
\end{equation}

With the derived mass function of the compact star $f_x(m)=0.268 M_\odot$, 
the binary mass ratio range $q=0.25-0.5$ obtained from our analysis allows the mass range
of the compact star $7.1-2.56 M_\odot$. 
The lower mass limit for $q=0.5$ formally admits a neutron star as the compact object in SS433. 
However, the optimal value $q\simeq 0.3$ independently obtained 
from the analysis of hard X-ray eclipses and precessional variability 
fully agrees with the mass ratio value $q=0.35$ found by \cite{hillwig08}
from optical spectroscopy. So we conclude that the compact object in SS433 is highly unlikely to be a neutron star and is indeed a black hole.

The mass of the black hole in SS433 $m_x\simeq 5 M_\odot$ is strongly based on its mass function derived from the optical spectroscopy \citep{hillwig08}. 
The value of the mass function is very sensitive to the radial velocity semi-amplitude of the 
optical star $\sim K_v^3$. As we emphasized earlier \citep{Cher05}, the measured 
semi-amplitude can be affected by the X-ray heating effect of the optical star, since we 
ascribe the observed radial velocity amplitude to the center of mass of the optical star 
filling its Roche lobe. The effect decreases with the binary mass ratio $q$. Although 
no signatures of the X-ray illumination effect on the radial velocity amplitude was
found by \cite{hillwig08}, the obtained semi-amplitude $K_v$ and hence the derived 
mass function of the compact star in SS433 should be treated with some reservation. 
Clearly, new high-resolution optical spectroscopy of SS433 is in order to accurately
measure the radial velocity curve of the optical star.

\section{Conclusions}
\label{s_concl}

The discovery of strong variability of the hard X-ray eclipse at precessional phases corresponding to the
maximum disk opening angle for the observer and regular 
significant precessional variability are 
the main findings of our INTEGRAL observations of SS433. 
It is shown that the duration of the primary X-ray eclipse at  
different epochs changes by $\sim 2$ times, the amplitude of the precessional variability is 5-7 times.  This implies that 
X-ray eclipses in SS433 are not purely geometrical, they are shaped
by absorption in a powerful stellar wind and gas streams from the optical star, as well as in the wind-wind collision region. 
So to infer the binary mass ratio $q$ from the 
analysis of X-ray eclipses, we have used only the ingress to
and upper envelope 
of egress out of the primary eclipse. The joint analysis of eclipsing light curve 
and the regular precessional light curve for off-eclipse 
flux from the system yielded the 
mass ratio estimate $q\simeq 0.25-0.5$. The relatively high value of the binary 
mass ratio $m_x/m_v$ (and hence the relatively large size of the Roche lobe
of the compact star) suggests an explanation to peculiarities of the optical
variability of SS433, in particular, to the observed substantial 
precessional variability of the minimum brightness at the middle
of the primary optical eclipse. 

Using the mass function of the optical star found by \cite{hillwig08} $f_v(m)=0.268 M_\odot$ 
and the value of $q=m_x/m_v\simeq 0.3$ 
corresponding to the absolute minimum of $\chi^2_{tot}$ residuals, we concluded that 
the masses of binary components of SS433 are $m_x\simeq 5.3 M_\odot$, 
$m_v\simeq 17.7 M_\odot$. The high mass of the compact  
object leaves no doubts that it is a black hole. 

The independence of the hard X-ray spectrum on the accretion disk
precessional phase suggests that the hard X-ray emission ($kT=20-100$~keV)
is formed in an extended, hot, quasi-isothermal corona. The heating
of the corona can be related to transformation of the kinetic energy of relativistic jets to an inhomogeneous wind outflow from 
the precessing supercritical accretion disk \citep{Begelman06}.

The Monte-Carlo simulations of broadband X-ray spectrum of SS433
at the maximum disk opening precessional phases \citep{Krivosheev08}
allowed us to determine the main physical characteristics of the 
hot corona (temperature $T_{cor}\simeq 20$~keV, Thomson optical depth  
$\tau\simeq 0.2$), as well as to estimate the mass outflow  rate in jets $\dot M_j=3\times 10^{19}$~g~s$^{-1}$ yielding the kinetic power of the jets $\sim 10^{39}$~erg~s$^{-1}$.

\section*{Acknowledgments}
The authors thank Dr. M. Revnivtsev for useful
discussions and the anonymous referee for constructive comments. The work is 
partially supported by the RFBR grants 07-02-00961 and 08-02-01220.


\begin{thebibliography}{99}

\bibitem[\protect\citeauthoryear{Antokhina \& Cherepashchuk}{1987}]{AntCher87}
Antokhina E.A., Cherepashchuk A.M., 1987, SvA, 31, 295

\bibitem[\protect\citeauthoryear{Antokhina et al.}{1992}]{Ant92}
Antokhina E.A., Seifina E.V., 
Cherepashchuk A.M., 1992, SvA, 36, 143

\bibitem[\protect\citeauthoryear{Begelman et al.}{2006}]{Begelman06}
Begelman M.C., King A.R., Pringle J.E., 2006, MNRAS, 370, 399

\bibitem[\protect\citeauthoryear{Brinkman et al.}{1989}]{brinkman89}
Brinkman W., Kawai N., Matsuoka M., 1989, Astron. Astrophys., 218, L13
%
\bibitem[\protect\citeauthoryear{Cherepashchuk}{1981}]{Cher81}
Cherepashchuk A.M., 1981, MNRAS, 194, 761

\bibitem[\protect\citeauthoryear{Cherepashchuk}{1989}]{Cher88}
Cherepashchuk A.M., 1989, Sov. Sci. Rev. Ap. Space Phys., Ed. by R.A.Sunyaev, v.7, p.185

\bibitem[\protect\citeauthoryear{Cherepashchuk et al.}{1995}]{Cher95}
Cherepashchuk A.M., Bychkov K.V., Seifina E.V., 
1995, ApSS, 229, 33

\bibitem[\protect\citeauthoryear{Cherepashchuk et al.}{2003}]{Cher03}
Cherepashchuk A.M., Sunyaev R.A., Seifina E.V., 
Panchenko I.E., Molkov S.V., Postnov K.A., 2003, AA, 411, 441

\bibitem[\protect\citeauthoryear{Cherepashchuk et al.}{2005}]{Cher05}
Cherepashchuk A.M., Sunyaev R.A., Fabrika S.N. et al, 2005, AA, 437, 561

\bibitem[\protect\citeauthoryear{Cherepashchuk et al.}{2007}]{Cher06}
Cherepashchuk A.M., Sunyaev R.A., Seifina E.V.  et al, 2007, Proc. 6th INTEGRAL Workshop,
eds. S. Grebenev, R.Sunyaev, C. Winkler, ESA SP-622, p. 319

\bibitem[\protect\citeauthoryear{Davydov et al.}{2008}]{Davydov08}
Davydov V.V., Esipov V.F., Cherepashchuk A.M., 2008, Astron. Rep., 52, 487

\bibitem[\protect\citeauthoryear{Fabrika}{2004}]{Fab04} Fabrika, S.N., 2004, Astrophys. Space Phys. Rev., 12, 1

\bibitem[\protect\citeauthoryear{Filippova et al.}{2006}]{Fil06}
Filippova E.V., Revnivtsev M., Fabrika S., Postnov K., Seifina E., 2006, AA, 460, 125

\bibitem[\protect\citeauthoryear{Gies et al.}{2002}]{Gies02}
Gies D.R., Huang W., McSwain M.V., 2002, ApJ, 578, L67

\bibitem[\protect\citeauthoryear{Goranskij et al.} {1998}]{Goransk98}
Goranskij V.P., Esipov V.F., Cherepashchuk A.M., 1998, Astron. Rep., 42, 209; ibid., p. 336

\bibitem[\protect\citeauthoryear{Hillwig et al.} {2004}]{Hillwig_ea04}
Hillwig T.C., Gies D.R., Huang W., McSwain M.V., Stark M.A.,
van der Meer A., Kaper L., 2004, ApJ, 615, 422

\bibitem[\protect\citeauthoryear{Hillwig \& Gies}{2008}]{hillwig08}
Hillwig T.C., Gies D.R., 2008, ApJ, 676, L37

\bibitem[\protect\citeauthoryear{Kawai et al.}{1989}]{Kawai89}
Kawai N., Matsuoka M., Pan H.-C., Stewart G.C., 
1989, PASJ, 41, 491

\bibitem[\protect\citeauthoryear{Kotani}{1998}]{KotaniPhD}
Kotani T., 1998, PhD. The Institute of Space and Astronautical Science,
Japan

\bibitem[\protect\citeauthoryear{Kotani et al.}{1998}]{Kotani98}
Kotani T., Kawai N., Matsuoka M., Brinkmann W., 1998, 
in Proc. IAU Symp. 188, eds. K. Koyama, S. Kitamoto, M. Itoh,
Dordrecht: Kluwer, p. 358


\bibitem[\protect\citeauthoryear{Krivosheyev et al.}{2009}]{Krivosheev08}
Krivosheyev Yu. M., Bisnovatyi-Kogan G.S., Cherepashchuk A.M., Postnov K.A., 2009, MNRAS, 394, 1674

\bibitem[\protect\citeauthoryear{Margon}{1984}]{Margon84}
Margon B., 1984, ARAA, 22, 507

\bibitem[\protect\citeauthoryear{Margon \& Anderson}{1989}]{MargonAnderson89}
Margon B., Anderson S.F., 1989, ApJ, 347, 507

\bibitem[\protect\citeauthoryear{Molkov et al.}{2004}]{Mol04}
Molkov S., Cherepashchuk A.M., Lutovinov A.A., Revnivtsev M.G., 
Postnov K.A., Sunyaev R.A., 2004, Astron. Lett., 30, 534

\bibitem[\protect\citeauthoryear{Namiki et al.}{2003}]{Namiki03}
Namiki M., Kawai N., Kotani T., Makishima K., 2003, PASJ, 55, 281

\bibitem[\protect\citeauthoryear{Revnivtsev et al.}{2004}]{Revnivtsev04}
Revnivtsev M., Burenin R., Fabrika S. 
et al., 2004, AA, 424, L5

\bibitem[\protect\citeauthoryear{Revnivtsev et al.}{2006}]{Revnivtsev06}
Revnivtsev M., Fabrika S., Abolmasov P.  
et al., 2006, AA, 447, 545

\bibitem[\protect\citeauthoryear{Shakura \& Sunyaev}{1973}]{ShS73}
Shakura N.I., Sunyaev R.A., 1973, AA, 24, 337


\end{thebibliography}
\end{document}